\documentclass[10pt,reqno]{amsart}
\usepackage{amsmath,amssymb,amsfonts,amsthm}
\usepackage[mathscr]{eucal}
\usepackage[russian,english]{babel}
\usepackage{hyperref}
\usepackage[utf8]{inputenc}
\usepackage[english]{babel}
\usepackage{color}

\title
{Higher derivative extensions of $3d$ Chern-Simons models:
conservation laws and stability}
\author{D.~S.~Kaparulin, I.~Yu.~Karataeva, S.~L.~Lyakhovich}
\address{Department of Quantum Field Theory, Tomsk State University, Tomsk 634050, Russia}
\email{dsc@phys.tsu.ru, karin@phys.tsu.ru, sll@phys.tsu.ru}

\begin{document}

\begin{abstract} We consider the class of  higher derivative $3d$
vector field models with the field equation operator being a
polynomial of the Chern-Simons operator. For $n$-th order theory of
this type, we provide a general receipt for constructing
$n$-parameter family of conserved second rank tensors. The family
includes the canonical energy-momentum tensor, which is unbounded,
while there are bounded conserved tensors that provide classical
stability of the system for certain combinations of the parameters
in the Lagrangian. We also demonstrate the examples of consistent
interactions which are compatible with the requirement of stability.
\end{abstract}

\maketitle

\section*{Introduction}

In this paper we consider a class of 1-form field $A=A_\mu dx^\mu$
models on $3d$ Minkowski space with the action
\begin{equation}\label{S}
    S=\frac{m^2}{2}\int  A \wedge(-a_0\ast A+a_1\frac{2}{m}dA+a_2\frac{4}{m^2}d\ast d A+ a_3\frac{8}{m^{3}}d\ast
    d\ast dA+  a_4\frac{16}{m^{4}} d\ast d\ast
    d\ast dA+  \ldots) \, ,
\end{equation}
where $m$ is a constant with dimension of mass, $a_0, a_1, a_2,
a_3,\ldots$ are some real dimensionless coefficients,  $*$ is Hodge
conjugation, and the signature is $(+,-,-)$. The coefficient $a_0
m^2$ corresponds to the usual mass term, $a_1 m$ is the Chern-Simons
mass, $a_2$ is a coefficient at Maxwell's Lagrangian, $a_3$
corresponds to the extended Chern-Simons Lagrangian \cite{DJ1999}
and the fourth order term appears in the Podolsky's electrodynamics
Lagrangian \cite{PED}. With appropriate choice of the coefficients
$a_k, k=0,1,2, ...$, this action reproduces various known $3d$
models, including the Chern-Simons-Proca \cite{TPN1984,DJ1984},
Maxwell-Chern-Simons \cite{DJT1982,DJT-PRL1982},
Maxwell-Chern-Simons-Proca \cite{BCS2001,DT2002} and the other
previously studied higher derivative models \cite{DR1988,SDN1999}.

In any dimension, inclusion of the higher derivative terms results
in the unbounded canonical energy, so classical stability becomes
the issue. It is also known that the ghost poles can emerge in the
propagator once higher derivatives are included in the action.

The specifics of higher order terms in three dimensions is that they
can be viewed as derived from the Chern-Simons term by the repeated
shift of the field by its strength: $A\mapsto A+2m^{-1}*dA$. As a
result, the operator of field equations is a polynomial in first
order operator $W=2m^{-1}*dA$. This special structure allows us to
make some conclusions concerning conservation laws and stability.
The observation is that the $n$-th order theory of the class
(\ref{S}) admits $n$ parametric family of conserved second rank
tensors whenever $a_0\neq 0$. Once $a_0=0$ (the theory is gauge
invariant in this case), there exists an $n-1$ parametric family of
conserved tensors. The canonical energy-momentum is included in the
family in every instance. We provide the general receipt for
constructing these conservation laws, and related symmetries. The
construction in fact applies to any system (of any field A, not
necessarily 1-form) with the operator of field equations being a
polynomial in another operator
\begin{equation}\label{Wn}
 M A=0, \qquad M= m^2\sum_{k=0}^na_kW^k\,,
\end{equation}
where $W$ can be any self-adjoint\footnote{The conjugation rule is
explained in the next section. } differential operator, $a_k$ are
real constants, and $a_n\neq 0$. We term the models of the type
(\ref{Wn})  \textit{derived} from the theory with equations $WA=0$.
For the case (\ref{S}), when $W=2m^{-1}*d$, we apply the general
procedure to explicitly deduce the conserved tensors for the third
order actions of this class. As we see, the bounded conserved
tensors are contained in the family, once the polynomial
$M(2m^{-1}*d)$ has only simple roots, or at most one double zero
root. In this generic case, the theory is classically stable even
though the canonical energy-momentum is unbounded. As we explain,
these models can admit certain interactions such that the stability
survives at nonlinear level. The case of multiple roots is special.
It also admits a family of $n$ conserved tensors, including the
canonical energy-momentum, though there are no bounded conserved
quantities in this family. As we see, the corresponding
representations of the Poincar\'e group are non-unitary, while in
the generic case, the representation decomposes into unitary ones.

The article is organized in the following way. In the next section
we describe the general structure of field equations in the higher
derivative models that fall into the class of derived theories
(\ref{Wn}). For the generic derived system of order $n$ we suggest a
procedure of constructing $n$-parametric family of conserved tensors
whose structure depends on the coefficients $a_k$ in the field
equations (\ref{Wn}). In Section 2, we explicitly construct the
families of conserved tensors for the theory (\ref{S}) involving
terms up to third order. As we see, four different cases are
possible from the viewpoint of existence the bounded representative
in the family of conserved quantities. These cases are distinguished
by the structure of roots in the polynomial (\ref{Wn}). Once the
positive conserved quantity exists, the theory is stable at
classical level, even though the canonical energy is unbounded. In
Section 3, we demonstrate the example of the self-interaction such
that the nonlinear theory remains stable. In conclusion, we
summarize the results and comment on stability of the theory
(\ref{S}) at quantum level.

\section{Derived theories, higher symmetries and conservation laws}\label{Sec2}
In this section, we consider the field equations of general
structure (\ref{Wn}). We demonstrate that combining the space-time
translations with the powers of operator $W$, one can construct
non-trivial higher order symmetries and find related conserved
tensors. The construction is quite general, it applies to any system
of the form (\ref{Wn}). The explicit details for the extension
Chern-Simons theory (\ref{S}) are provided in Section 2.

\subsection{Derived theories.}
Consider a set of fields $A^J (x)$ on $d-$dimensional Minkowski
space with local coordinates $x^\mu$. The multi-index $J$
accommodates all the tensor, spinor, isotopic indices labeling the
field components. Here, we suppose that the theory admits
appropriate constant metrics that can be used to rise and lower the
multi-indices. In this setting, any local linear system of field
equations can be represented in the following form:
\begin{equation}\label{M}
    M_{IJ}(\partial)A^J=0 \, ,
\end{equation}
where $M_{IJ}(\partial)$ is a square matrix whose entries are
polynomials in the formal variables $\partial_\mu$. If
$\partial_\mu$ are understood as the partial derivatives in
Minkowski space coordinates $x^\mu$, (\ref{M}) will be a linear PDE
system. The formal adjoint to the operator $M$ is defined by
\begin{equation}\label{HCon}
    M^\dagger_{IJ}(\partial)= M_{JI}(-\partial ).
\end{equation}
The field equations (\ref{M}) are variational whenever
$M=M^\dagger$, in which case the action reads
\begin{equation}\label{LMI}
S=\int d^dx L\,,\qquad L=\frac{1}{2}A^I M_{IJ}(\partial)A^J\, .
\end{equation}
Let us further suppose  that the self-adjoint linear differential
operator $W_{IJ}(\partial)$ exists (Cf.(\ref{Wn})) such that the
operator of field equations is polynomial in $W$:
\begin{equation}\label{MExp}
M(W)=m^2\sum_{k=0}^n a_k
W^k=a_nm^2\prod_{i=1}^r(W-\lambda_i)^{p_i}\prod_{j=1}^s(W^2-(\omega_j+\overline{\omega}_j)W+\omega_j\overline\omega_j)^{q_j}\,.
\end{equation}
The real numbers $\lambda_i$ and complex conjugate numbers
$\omega_j,\overline{\omega}_j$ are the roots of the polynomial
$M(W)$ with multiplicities $p_i$ and $q_j$, respectively. The
multiplicity of roots is connected with the total degree of the
polynomial
$$
\sum_{i=1}^r p_i+2\sum_{j=1}^s q_j=n\,.
$$
If $W$ is a differential operator of finite order $n_W$, the order
of PDE system (\ref{M}) will not exceed $n\times n_W$.

Once the field equation operator $M(\partial)$ is a polynomial of
another self-adjoint operator $W(\partial)$, we say that the theory
is a derived model. In \cite{KLS2014EurPhysC}, the special case of
the factorization (\ref{MExp}) was studied, where $M$ has two
different simple real roots in $W$. This simple assumption has
far-reaching consequences. In particular, each of the factors
defines its own Lagrangian theory whose order is lower than that of
the derived theory. Let us mention some of these consequences
noticed in \cite{KLS2014EurPhysC}. Once the two lower order theories
are translation invariant, the derived higher derivative theory has
a two-parameter family of independent conserved tensors. This family
includes the canonical energy-momentum tensor of the derived theory.
The canonical energy is unbounded in general, as it should be in the
higher derivative system, while some other conserved quantities can
be bounded in this family. The existence of the bounded conserved
quantities guaranties the classical stability of dynamics. As it was
demonstrated in the paper \cite{KLS2014EurPhysC}, every conserved
tensor in this family can be connected to translation invariance of
the system by appropriate Lagrange anchor\footnote{The notion of the
Lagrange anchor was introduced in the work \cite{KLS-JHEP05} in
relation to the path-integral quantization of not necessarily
Lagrangian systems. Later it was shown that every Lagrange anchor
admitted by the equations of motion maps the conserved quantity to
the symmetry of equations \cite{KLS-JMP10}. In the paper
\cite{KLS2014EurPhysC} it was noticed that once the operator $M$
decomposes into two self-adjoint independent factors, the equations
admit a two-parameter family of Lagrange anchors such that map any
representative of the family of conserved tensors to the space-time
translation. In this sense, any of these tensors can be understood
as energy-momentum of the theory.}. As we will see in this section,
for any derived system (\ref{Wn}), one can construct $n$-parameter
family of conserved tensors, where $n$ is the order of polynomial
$M(W)$.

We consider two ways of constructing conserved tensors in the
derived theories.  At first, we make notice that the symmetry
algebra of the derived theory includes higher order symmetries
generated by the operator $W$, and translations, once $W$ is
translation invariant. Then, we derive the conserved tensors from
these symmetries by the Noether theorem. Another option employs the
procedure of reducing the order of higher derivative theory
(\ref{Wn}) by assigning a lower order system to every irreducible
factor in decomposition of the polynomial (\ref{MExp}). Then, making
use of the canonical conserved tensors for the lower order systems,
we get the family of the conserved tensors for the original theory
(\ref{Wn}). Although the Noether theorem provides a uniform way for
deducing conservation laws from given symmetries, the conserved
tensors obtained from the lower order equivalent system appear in a
more convenient form in this case, and we will use them for further
analysis of stability.

\subsection{Higher order symmetries and conservation laws.} Provided the
operator $W$ is translation invariant, the action (\ref{LMI}) admits
the following symmetry transformations:
\begin{equation}\label{Sym0}
\delta_{\varepsilon}A^J=-\varepsilon^\alpha \partial_\alpha
(W^kA)^J\,,\qquad k=0,\ldots,n-1.
\end{equation}
The space-time translations correspond to $k=0$. The higher order
transformations with $k=n,n+1,\ldots$ are equivalent to the lower
order ones with account of the equations of motion (\ref{M}), while
for $k<n$ one has independent symmetries. By the Noether theorem one
can link the symmetries (\ref{Sym0}) with the conserved tensors
\begin{equation}\label{CT}
(T^{k}){}^\mu_{\phantom{\mu}\nu}(A)\,,\qquad \partial_\mu
(T^{k})^\mu_{\phantom{\mu}\nu}=-\left(\partial_\nu
(W^kA)^J\right)(MA)_J\,,\qquad k=0,1,\ldots,n-1\,.
\end{equation}
Here, $k=0$ corresponds to the usual energy-momentum tensor. There
are $n$ independent tensors in the set (\ref{CT}).

\subsection{Conservation laws by the reduction of order.}
Consider the polynomial (\ref{MExp}). Denote the cofactors to the
real roots  $\lambda_i$ and complex roots $\omega_j$  by $\Lambda_i$
and $\Omega_j$, respectively,
\begin{equation}\label{LObar}\begin{array}{l}\displaystyle
\Lambda_i=\prod_{k\neq
i}(W-\lambda_k)^{p_k}\prod_{j=1}^s(W^2-(\omega_j+\overline{\omega}_j)W+\omega_j\overline\omega_j)^{q_j}
\, ,\\[3mm]\displaystyle \Omega_j=\prod_{i=1}^r(W-\lambda_i)^{p_i}\prod_{k\neq
j}(W^2-(\omega_k+\overline{\omega}_k)W+\omega_k\overline\omega_k)^{q_k}\,.
\end{array}\end{equation}
By definition, the polynomials $\Lambda_i(W)$ and $\Omega_j(W)$ are
coprime. Obviously,
$$
M=a_nm^2(W-\lambda)^{p_i}\Lambda_i=a_nm^2(W^2-(\omega_j+\bar{\omega}_j)W+\omega_j\overline\omega_j)^{q_j}\Omega_j\qquad
(\text{no summation in } i,j)\,.
$$

For each cofactor, we introduce the new set of fields,
\begin{equation}\label{xzA}
(\xi_i)^J=(\Lambda_i A)^J\,,\qquad
i=1,\ldots,r,\qquad(\zeta_j)^J=(\Omega_j A)^J\,,\qquad j=1,\ldots,s,
\end{equation}
called \emph{components}. Once the original fields $A$ are subject
to the original field equations (\ref{M}), the components satisfy
the lower order derived equations
\begin{equation}\label{xzEq}
a_nm^2(W-\lambda_i)^{p_i}\xi_i=0\,,\qquad
a_nm^2(W^2-(\omega_j+\overline\omega_j)W+\omega_j\overline{\omega}_j)^{q_j}\zeta_j=0\,,
\end{equation}
where $p_i, q_j$ are the multiplicities of the roots $\lambda_i,
\omega_j$ in the operator of the original equations (\ref{MExp}).

It is easy to see the one-to-one correspondence between solutions of
these equations and the original system (\ref{M}). The inverse
transformation to (\ref{xzA}) is established by the relations
\begin{equation}\label{Axz}
A^J=\sum_{i=1}^r(B{}_i\xi{}_i)^J+
\sum_{j=1}^s(C{}_j\zeta{}_j)^J\,,\qquad
B_i=\sum_{p=0}^{p_i-1}b{}_i^p W^p\,,\qquad
C_j=\sum_{q=0}^{2q_j-1}c{}_j^q W^q\,,
\end{equation}
where the polynomials $B_i(W)$ and $C_j(W)$ can be found by the
method of undetermined coefficients. The coefficients $b_i^p,c_j^q$
are defined by the relation
\begin{equation}\label{BC=1}
\sum_{i=1}^r B_i\Lambda_i+\sum_{j=1}^s C_j\Omega_j=1\,.
\end{equation}
The last equality is just Bezout's identity for the coprime
univariate polynomials $\Lambda_i(W)$ and $\Omega_j(W)$.

Whenever the equivalent formulation (\ref{xzEq}) is known, the
conserved tensors can be obtained by applying the relation
(\ref{CT}) separately to every component and then summarizing the
results. We denote the conserved tensors for the components by
\begin{equation}\label{xzCT}
    (\tau^p_i)^\mu_{\phantom{\mu}\nu}(\xi_i)\,,\quad p=0,\ldots,p_i-1\,,\qquad
    (\sigma^q_j)^\mu_{\phantom{\mu}\nu}(\zeta_j)\,,\quad
    q=0,\ldots,2q_j-1\,,
\end{equation}
where the indices $i,j$ label the corresponding components
(\ref{xzA}) while $p_i, q_j$ are the multiplicities of corresponding
roots (\ref{MExp}). The conserved tensors of original derived theory
are obtained by substitution (\ref{xzA}):
\begin{equation}\label{2CT}
    (T^p_i)^\mu_{\phantom{\mu}\nu}(\Lambda_iA)=(\tau^p_i)^\mu_{\phantom{\mu}\nu}(\xi_i)\Big|_{\xi_i=\Lambda_iA}\,,
    \qquad(U^q_j)^\mu_{\phantom{\mu}\nu}(\Omega_jA)=(\sigma^q_j)^\mu_{\phantom{\mu}\nu}(\zeta_j)\Big|_{\zeta_j=\Omega_jA}\,.
\end{equation}
By construction,
\begin{equation}\label{div2CT}
    \partial_\mu(T^p_i)^\mu_{\phantom{\mu}\nu}(\Lambda_iA)=-(\partial_\nu(W^p\Lambda_i A)^J)(MA)_J\,,
    \qquad \partial_\mu(U^q_j)^\mu_{\phantom{\mu}\nu}=-(\partial_\nu(W^q\Omega_j A)^J)(MA)_J.
\end{equation}
There are $n$ conserved tensors (\ref{2CT}). The relationship
between ``new" and ``old" conserved tensors is established by
comparing their divergences (\ref{CT}) and (\ref{div2CT}). In
particular, the canonical energy-momentum tensor of the derived
theory (\ref{Wn}) has the following representation:
$$
(T^0)^\mu_{\phantom{\mu}\nu}(A)=\sum_{i=1}^r\sum_{p=0}^{p_i-1}
b{}^p_i
(T^p_i)^\mu_{\phantom{\mu}\nu}(\Lambda_iA)+\sum_{j=1}^r\sum_{q=0}^{2q_j-1}
c{}^q_j (U^q_j)^\mu_{\phantom{\mu}\nu}(\Omega_jA)\,,
$$
with the coefficients of linear combination being defined by Rel.
(\ref{BC=1}).

Notice that some combinations of the conserved tensors (\ref{CT}) or
(\ref{2CT}) may  be trivial. A conserved tensor is said to be
trivial if it is given by the divergence of an antisymmetric tensor
modulo equations of motion, i.e.,
$$
T^\mu_{\phantom{\mu}\nu}(A)\Big|_{MA=0}=\partial_{\alpha}\Sigma{}^{\alpha\mu}_{\phantom{\alpha\mu}\nu}\,,\qquad
\Sigma{}^{\alpha\mu}_{\phantom{\alpha\mu}\nu}=-\Sigma{}^{\mu\alpha}_{\phantom{\alpha\mu}\nu}\,.
$$
The trivial conserved tensors do not result in any conserved
quantity and have to be systematically ignored. However, we provide
the expressions for the conserved tensors modulo divergence terms,
but keep the contributions from the equations of motion. Consistency
of the computations can then be verified by taking the divergence,
see (\ref{CT}) and (\ref{div2CT}).

As the issue of stability is concerned, the \emph{positive}
conserved tensors are relevant. By positive tensor we mean the one
whose $00$-component is positive for any solution which is not a
pure gauge. We consider the ansatz for the general conserved tensor
of the derived theory (\ref{M}) in the form
\begin{equation}\label{WT00}
T^\mu_{\phantom{\mu}\nu}(A)=\sum_{i=1}^r\sum_{p=0}^{p_i-1}\beta^p_i(T^p_i)^\mu_{\phantom{\mu}\nu}(\Lambda_iA)+
\sum_{j=1}^s\sum_{q=0}^{2q_j-1}\gamma^q_j(U^q_j)^\mu_{\phantom{\mu}\nu}(\Omega_jA)\,
.
\end{equation}
The ansatz means that we consider the conserved tensors being
additive in the contributions from bilinear combinations of
$\Lambda_iA$ and $\Omega_jA$, where $\Lambda_i, \Omega_j$ are the
cofactors (\ref{LObar}) to the real roots $\lambda_i$ and complex
roots $\omega_j$ in the decomposition (\ref{MExp}). The quadratic
forms $(T^p_i)$ and $(U^q_j)$ are defined by Rels. (\ref{xzCT}),
(\ref{2CT}). In fact, they represent in terms of the original field
$A$ the conserved tensors (\ref{CT}) of the component fields
$\xi_i,\zeta_j$ subject to equations (\ref{xzEq}). Here, $(T^0_i),
(U^0_j)$ are just the energy-momentum tensors for the component
fields $\xi_i, \zeta_j$ expressed in terms of $A$ by substitution
(\ref{xzA}), while $p,q>0$ correspond to the higher order symmetries
(\ref{Sym0}) of component fields.

As far as the components (\ref{xzA}) are independent, the conserved
tensor (\ref{WT00}) is positive if and only if so are the tensors
$$
\sum_{p=0}^{p_i}\beta^p_i(T^p_i)^\mu_{\phantom{\mu}\nu}(\xi_i)\,,
\qquad\sum_{q=0}^{q_j-1}\gamma^q_j(U^q_j)^\mu_{\phantom{\mu}\nu}(\zeta_j)\,.
$$
In the other words, the derived theory (\ref{Wn}) is stable if and
only if all the components (\ref{xzA}) are stable.

Below, we examine the third order extension of the Chern-Simons
theory from the viewpoint of existence of bounded $00$-components of
conserved tensors we found above.

\section{Conserved tensors in the third order extension of the Chern-Simons theory}

The field equations of higher derivative extension of the
Chern-Simons model (\ref{S}) fall into the class of derived theories
(\ref{Wn}), with $W$ being composition of the Hodge and de Rham
operators:
\begin{equation}\label{WCS}
(W){}_\mu^{\phantom{\mu}\nu}=(2m^{-1}*d){}_\mu^{\phantom{\mu}\nu}\,,\qquad
W{}_\mu^{\phantom{\mu}\nu}A_\nu=m^{-1}
\varepsilon{}_{\mu}^{\phantom{\mu}\alpha\nu}\partial_\alpha
A_\nu\,,\qquad \varepsilon_{012}=\varepsilon^{012}=1\,.
\end{equation}

The $n$-th order theory (\ref{S}) has $n$ degrees of freedom if
there are no zero roots in the polynomial (\ref{MExp}). If the zero
root exists of any multiplicity (including simple zero root) one
degree of freedom is gauged out by transformation $\delta_\chi
A=d\chi(x)$, so the theory has $n-1$ DoF. The theory (\ref{S})
describes a (decomposable) representation of the proper Poincar\'e
group. Its indecomposable sub-representations are described by the
components (\ref{xzA}). In particular, the field content of the
theory with simple real roots includes $n$ massive vector fields
that satisfy the Chern-Simons-Proca equations ($n-1$ massive fields
and one Chern-Simons field in the gauge case).\footnote{The
irreducible massive vector corresponds to the massive representation
of the proper Poincar\'e group. Being subject to the self-duality
equation proposed in \cite{TPN1984,DJ1984}, it has one physical
polarization. On the generalities of the Poincar\'e group unitary
irreducible representations in $d=3$ we refer to
\cite{Binegar-JMP1982,Grigore1,Grigore2}.} Double zero root
describes Maxwell's theory. A pair of complex conjugate roots
results in the theory with tachyons. The representations related to
multiple nonzero roots and zero root of multiplicity higher than 2
are non-unitary. The case of multiple roots is special because the
set of conserved tensors (\ref{xzCT}) includes a number of terms
corresponding to the multiplicity of root. One of the terms
corresponds to the energy-momentum tensor of the component, while
the others are connected to the higher order symmetries of the
components
$$
\delta_\varepsilon\xi_i=-\varepsilon^\alpha\partial_\alpha
(W^p\xi_i), \,\qquad p=1, \ldots\,, p_i-1, \,\qquad
\delta_\varepsilon\zeta_j=-\varepsilon^\alpha\partial_\alpha
(W^q\zeta_j), \,\qquad q=1, \ldots\,, 2q_j-1,
$$ where $p_i, q_j$ are multiplicities of real
and complex roots. Below we will observe that equations do not have
positive conserved quantities in the family (\ref{WCT}) once they
involve tachyon or non-unitary representations (that corresponds to
complex, double or higher multiplicity nonzero real or triple or
higher multiplicity zero roots). The models leading to the unitary
representations (that corresponds to simple roots, or at most one
double zero root in (\ref{MExp})) admit the conserved tensors with
bounded $00$-component even though the canonical energy is unbounded
in all the instances.

The conserved tensors (\ref{CT}) and (\ref{2CT}) of higher
derivative extension of the Chern-Simons model are given by
\begin{equation}\label{WCT}\begin{array}{l}\displaystyle
(T^k){}^\mu_{\phantom{\mu}\nu}=-\frac{m^2}{2}\Big\{\frac{1}{m^2}\delta^\mu_{\phantom{\mu}\nu}(W^kA)^\alpha
(MA)_\alpha+\sum_{l=0}^n a_l
(t^{k,l})^\mu_{\phantom{\mu}\nu}(A)\Big\}\,,\\[3mm]\displaystyle
(T^p_i){}^\mu_{\phantom{\mu}\nu}=-\frac{m^2}{2}\Big\{\frac{1}{m^2}\delta^\mu_{\phantom{\mu}\nu}(W^pA)^\alpha
(MA)_\alpha+a_n\sum_{l=0}^{p_i}\frac{p_i!(-\lambda_i)^{p_i-l}}{l!(p_i-l)!}
(t^{p,l})^\mu_{\phantom{\mu}\nu}(\Lambda_i A)\Big\}\,,\\[3mm]\displaystyle
(U^q_j){}^\mu_{\phantom{\mu}\nu}=-\frac{m^2}{2}\Big\{\frac{1}{m^2}\delta^\mu_{\phantom{\mu}\nu}(W^qA)^\alpha
(MA)_\alpha+a_n\sum_{l=0}^{q_j}\sum_{k=0}^{q_j-l}\frac{q_j!(-\omega_j-\overline\omega_j)^{k}(\omega_j\overline\omega_j)^{q_j-l-k}}{l!k!(q_j-l-k)!}
(t^{q,2l+k})^\mu_{\phantom{\mu}\nu}(\Omega_j A)\Big\}\,,\\[3mm]\displaystyle
\qquad k=0,1,\ldots,n-1\,,\qquad p=0,1,\ldots,p_i-1\,,\qquad
q=0,1,\ldots,2q_j-1\,,
\end{array}\end{equation}
where the notation is used
$$
(t^{k,l})^\mu_{\phantom{\mu}\nu}=\frac{1}{m}\varepsilon^{\mu\alpha\beta}\Big[\sum_{s=1}^{l-k}(W^{k+s-1}A)_\alpha
    \partial_\nu (W^{l-s}A)_\beta-\sum_{s=1}^{k-l}(W^{k-s}A)_\alpha
    \partial_\nu (W^{l+s-1}A)_\beta\Big]\,,\qquad
    (t^{k,l})^\mu_{\phantom{\mu}\nu}\Big|_{l=k}=0\,.
$$
The expressions for the conserved tensors (\ref{WCT}) can be
simplified making use of the identity
\begin{equation}\label{EDelta}\begin{array}{l}\displaystyle
\frac{1}{m}\varepsilon^{\mu\alpha\beta} (W^k A)_\alpha \partial_\nu
(W^l A)_\beta=(W^{k+1}A)^\mu (W^l A)_\nu+ (W^{l+1}A)^\mu (W^k)_\nu-
\delta^\mu_{\phantom{\mu}\nu} (W^{l+1}A)^\alpha (W^k
A)_\alpha-\\[3mm]\displaystyle\qquad-\frac{1}{m}\partial_\alpha (\varepsilon^{\mu\alpha\beta} (W^l A)_\nu
(W^kA)_\beta)\,,\qquad k,l\geq 0\,.\end{array}\end{equation}
Applying this formula one can express all the conserved tensors in
terms of \mbox{$W^kA,k=0,\ldots,n-1$.}

For $a_0=0$, only $n-1$ of $n$ conserved tensors (\ref{WCT}) are
non-trivial. The trivial conserved tensor reads
\begin{equation}\label{triv}
(T^{p_i-1}_{i})^\mu_{\phantom{\mu}\nu}\equiv\sum_{k=p_i}^n
a_k(T^{k-1})^\mu_{\phantom{\mu}\nu}=-(MA)^\mu(W^{p_i-1}\Lambda_iA)_\nu+
\frac{m}{2}\partial_\alpha(\varepsilon^{\mu\alpha\beta}(W^{p_i-1}\Lambda_iA)_\nu(W^{p_i-1}\Lambda_iA)_\beta)\,,
\end{equation}
with $\lambda_i=0.$ The simplest example of that kind is provided by
the energy-momentum tensor for the Chern-Simons theory, where
$n=1\,,\lambda_1=0\,,p_1=1$.

With account of (\ref{triv}), we consider the following ansatz for
the general conserved tensor of the derived theory~(\ref{M}):
\begin{equation}\label{T00}
T^\mu_{\phantom{\mu}\nu}(W^{n-1}A,\ldots,WA,A)=\sum_{i=1}^r\sum_{p=0}^{\widetilde{p}_i-1}\beta^p_i(T^p_i)^\mu_{\phantom{\mu}\nu}(\Lambda_iA)+
\sum_{j=1}^s\sum_{q=0}^{2q_j-1}\gamma^q_j(U^q_j)^\mu_{\phantom{\mu}\nu}(\Omega_jA)\,,
\end{equation}
where $\widetilde{p}_i=p_i$ if $\lambda_i\neq0$ and
$\widetilde{p}_i=p_i-1$ otherwise. Here, its $00$-component is given
by the quadratic form in $W^kA,k=0,\ldots,n-1$ ( $k=1,\ldots,n-1$ in
case $a_0=0$). Identification of the range of the parameters $\beta$
and $\gamma$ that satisfy positivity condition is a well-known
problem of linear algebra. It can be always solved in various ways,
for example, by the Silvester criterion.

Let us turn to the case when $a_k=0$ for $k>3$ and $a_3=1$. This is
the most general case of the third order derived theory. The
equations of motion (\ref{Wn}) read
\begin{equation}\label{EoM3}\begin{array}{ll}\displaystyle
M_\mu^{\phantom{\mu}\nu} A_\nu=0\,,& \displaystyle
M_\mu^{\phantom{\mu}\nu}=
m^2(W_\mu^{\phantom{\mu}\alpha}W_\alpha^{\phantom{\mu}\beta}W_\beta^{\phantom{\mu}\nu}+
a_2W_\mu^{\phantom{\mu}\beta}W_\beta^{\phantom{\mu}\nu}+
a_1W_\mu^{\phantom{\mu}\nu}+a_0\delta_\mu^{\phantom{\alpha}\nu})=\\[3mm]\displaystyle
&\displaystyle\phantom{M^\mu_{\phantom{\mu}\nu}}=-\frac{1}{m}\Box\varepsilon_{\mu}^{\phantom{\mu}\alpha\nu}\partial_\alpha-
a_2(\Box\delta_\mu^{\phantom{\alpha}\nu}-\partial_\mu\partial^\nu)+
a_1m\varepsilon_{\mu}^{\phantom{\alpha}\alpha\nu}
\partial_\alpha+a_0m^2\delta_\mu^{\phantom{\alpha}\nu}\,.
\end{array}\end{equation} This model has a three-parameter family of
conserved tensors if $a_0\neq0$ and a two-parameter family if
$a_0=0$. Depending on the structure of roots in the decomposition
(\ref{MExp}) for the third order equations (\ref{EoM3}), the four
different cases are seen with different behavior of $00$-component
of the conserved tensors:

Case A: Three different real roots. The family of conserved tensors
includes the one with  positive $00$-component.

Case B: Simple real root and real root of multiplicity 2. The
conserved tensor exists with the positive $00$-component if the
double root is zero, otherwise the conserved quantity is unbounded.

Case C: Simple real root and pair of complex conjugate roots. The
conserved tensor with the positive $00$-component does not exist.

Case D: Real root of multiplicity 3.  The conserved tensor with the
positive $00$-component does not exist.

Below we elaborate on each case separately.

\subsection{Case A} The coefficients $a_0,a_1,a_2$ are
defined by three real roots $\lambda_1<\lambda_2<\lambda_3$ of the
polynomial (\ref{MExp}),
$$
a_2=-(\lambda_1+\lambda_2+\lambda_3)\,,\qquad
a_1=\lambda_1\lambda_2+\lambda_2\lambda_3+\lambda_1\lambda_3\,,\qquad
a_0=-\lambda_1\lambda_2\lambda_3\,.
$$
The factorization (\ref{MExp}) for the equations of motion (\ref{M})
reads
$$
M=m^2\prod_{i=1}^{3}(W-\lambda_i)\,,
$$
that corresponds to  $r=3$, $s=0$, $p_i=1$.

The general solution to the theory (\ref{EoM3}) is decomposed into
three components (\ref{xzA}),
\begin{equation}\label{xi3}
\xi_i=\Lambda_iA\,,\qquad \Lambda_i=\prod_{j\neq
i}(W-\lambda_j)\,,\qquad i=1,2,3\,,
\end{equation}
that satisfy the  Chern-Simons-Proca equations
\begin{equation}\label{xi3Eq}
m^2(W-\lambda_i)\xi_i=0\,.
\end{equation}
Each of the equations describes the massive vector field with the
mass $m|\lambda_i|$. Thus, the third-order theory describes a
collection of three massive fields with different masses. At the
level of propagator, the decomposition into irreducible components
has been noticed in already in the original paper \cite{DJ1999},
where the third order extension was proposed for the Cher-Simons
theory. In this paper we see the decomposition at the level of
solutions to the equations of motion and elaborate on conserved
tensors. In case of second-order theory, $n=2$, the decomposition
into components was noticed \cite{PK1985}. The solution (\ref{Axz})
to the original theory (\ref{EoM3}) is reconstructed by the formula
\begin{equation}\label{B3}
A=\sum_{i=1}^3 B_i\xi_i\,,\qquad B_{i}\equiv b^0_i=\prod_{j\neq
i}(\lambda_i-\lambda_j)^{-1}\,.
\end{equation}
The conserved tensors (\ref{xzCT}) are labelled by the indices
$i=1,2,3$, $p=0$ and have the form
\begin{equation}\label{3T}
(T^0_i)^\mu_{\phantom{\mu}\nu}(\Lambda_iA)=-\frac{m^2}{2}\Big\{2\lambda_i(\Lambda_iA)^\mu(\Lambda_iA)_\nu-
\lambda_i\delta^\mu_{\phantom{\mu}\nu}(\Lambda_iA)^\alpha(\Lambda_iA)_\alpha\Big\}-(MA)^\mu
(\Lambda_iA)_\nu\,.
\end{equation}
The sign of the corresponding $00$-component coincides with the sign
of $-\lambda_i$,
\begin{equation}\label{3T00}
(T^0_i)^0_{\phantom{0}0}(\Lambda_iA)=-\frac{m^2}{2}\lambda_i(\Lambda_iA\,,\Lambda_i
A)-(MA)^0 (\Lambda_iA)_0\,.
\end{equation}
Here, the Euclidean scalar product is used,
$$
(\Lambda_iA\,,\Lambda_i A)=((\Lambda_i A)_0)^2+((\Lambda_i
A)_1)^2+((\Lambda_i A)_2)^2>0\,.
$$

The conserved tensors (\ref{3T}) can be combined into the tensor
\begin{equation}\label{LS3T}
T^\mu_{\phantom{\mu}\nu}(A)=
\sum_{i=1}^3\beta^0_i(T^0_i)^\mu_{\phantom{\mu}\nu}(\Lambda_iA)\,
\end{equation}
with the positive $00$-component if and only if
$-\beta{}^0_i\lambda{}_i>0$. This result admits simple physical
interpretation. Each of the tensors (\ref{3T}) has the sense of the
energy-momentum tensor of the component $\xi_i$. The $00$-component
of the general conserved tensor (\ref{LS3T}) is bounded if the
contributions of all the components have the same sign. In contrast,
the $00$-component of the canonical energy-momentum tensor with
$\beta^0_i=b^0_i$ is always unbounded because the different
components contribute with different signs.

Finally, there is an option when one of the roots is zero. In this
case, the corresponding conserved tensor becomes trivial and the
positivity of the $00$-component of the general conserved tensor
(\ref{T00}) is ensured by imposing condition
$-\beta{}^0_i\lambda{}_i>0$ for the nonzero roots. The
$00$-component of the canonical energy-momentum tensor is again
unbounded.

\subsection{Cases B and C} We deduce the explicit expressions for the conserved quantities in the Case C.
Corresponding expressions for Case B follow from the ones of the
Case C by setting the imaginary part of complex root to zero.

The polynomial (\ref{MExp}) has the simple real root $\lambda_1$ and
the simple complex root $\omega_1$, i.e.,
$$
M=m^2(W-\lambda_1)(W^2-(\omega_1+\overline{\omega}_1)W+\omega_1\overline\omega_1)\,.
$$
Here, $r=1$ and $s=1$, so the indices $i,j$ numerating real and
complex roots take a single value $i=j=1$. The parametrization for
the coefficients $a_0,a_1,a_2$ of the polynomial (\ref{Wn}) reads
$$
a_2=-(\lambda_1+\omega_1+\overline\omega_1)\,,\qquad
a_1=\lambda_1(\omega_1+\overline\omega_1)+\omega_1\overline\omega_1\,,\qquad
a_0=-\lambda_1\omega_1\overline\omega_1\,.
$$

The general solution to the theory (\ref{EoM3}) decomposes into the
pair of components (\ref{xzA})
$$
\xi_1= \Lambda_1 A\,,\qquad\zeta_1=\Omega_1 A\,,\qquad
\Lambda_1=W^2-(\omega_1+\overline{\omega}_1)W+\omega_1\overline\omega_1\,,\qquad
\Omega_1=W-\lambda_1
$$
that satisfy the first-order and the second-order equations
(\ref{xzEq}),
\begin{equation}\label{xizetaEq}
m^2(W-\lambda_1)\xi_1=0\,,\qquad
m^2(W^2-(\omega_1+\overline{\omega}_1)W+\omega_1\overline\omega_1)\zeta_1=0\,,
\end{equation}
respectively. The equations for the $\xi$-component correspond to
the Chern-Simons-Proca theory \cite{TPN1984, DJ1984} with mass
$m|\lambda_1|$ . The $\zeta-$field satisfies the (tachyon)
Maxwell-Chern-Simons-Proca equations \cite{BCS2001,DT2002}. The
solution (\ref{Axz}) to the original theory (\ref{EoM3}) is
reconstructed by the formula
$$
A=\frac{1}{(\lambda_1-\omega_1)(\lambda_1-\overline\omega_1)}\xi_1+
\Big[\frac{W-\omega_1}{(\overline\omega_1-\lambda_1)(\overline\omega_1-\omega_1)}+
\frac{W-\overline\omega_1}{(\omega_1-\lambda_1)(\omega_1-\overline\omega_1)}\Big]\zeta_1\,.
$$
The conserved tensors (\ref{WCT}) of the theory are parameterized by
the indices $p=0$ and $q=0,1$. The expressions for the tensors have
the form
\begin{equation}\label{TRC}\begin{array}{l}\displaystyle
(T^0_1)^\mu_{\phantom{\mu}\nu}=-\frac{m^2}{2}\Big\{2\lambda_1(\Lambda_1
A)^\mu(\Lambda_1
A)_\nu-\lambda_1\delta^\mu_{\phantom{\mu}\nu}(\Lambda_1
A)^\alpha(\Lambda_1 A)_\alpha\Big\}-(MA)^\mu (\Lambda_1A)_\nu\,,\\[3mm]\displaystyle
(U^0_1)^\mu_{\phantom{\mu}\nu}=-\frac{m^2}{2}\Big\{2(W\Omega_1
A)^\mu(W\Omega_1 A)_\nu-2\omega_1\overline\omega_1(\Omega_1
A)^\mu(\Omega_1 A)_\nu-\delta^\mu_{\phantom{\mu}\nu} \big[(W\Omega_1
A)^\alpha(W\Omega_1 A)_\alpha-\\[3mm]\displaystyle\qquad-\omega_1\overline\omega_1(\Omega_1
A)^\alpha(\Omega_1 A)_\alpha\big]\Big\}-(MA)^\mu (\Omega_1A)_\nu\,,\\[3mm]\displaystyle
(U^1_1)^\mu_{\phantom{\mu}\nu}=-\frac{m^2}{2}\Big\{
2(\omega_1+\overline\omega_1) (W\Omega_1 A)^\mu(W\Omega_1
A)_\nu-2\omega_1\overline\omega_1\big((\Omega_1 A)^\mu(W\Omega_1
A)_\nu+(W\Omega_1 A)^\mu(\Omega_1
A)_\nu\big)-\\[3mm]\displaystyle\qquad-\delta^\mu_{\phantom{\mu}\nu}\big[(\omega_1+\overline\omega_1)(W\Omega_1
A)^\alpha(W\Omega_1 A)_\alpha-2\omega_1\overline\omega_1(W\Omega_1
A)^\alpha(\Omega_1 A)_\alpha\big]\Big\}-(MA)^\mu (W\Omega_1A)_\nu\,.
\end{array}\end{equation}
The $00$-components read
\begin{equation}\label{TRC00}\begin{array}{l}\displaystyle
(T^0_1)^0_{\phantom{\mu}0}=-\frac{m^2}{2}\lambda_1(\Lambda_1 A\,,\Lambda_1 A)-(MA)^0 (\Lambda_1A)_0\,,\\[3mm]\displaystyle
(U^0_1)^0_{\phantom{\mu}0}=-\frac{m^2}{2}\Big\{(W\Omega_1 A\,,
W\Omega_1 A)-
\omega_1\overline\omega_1(\Omega_1 A\,,\Omega_1 A)\Big\}-(MA)^0 (\Omega_1A)_0\,,\\[3mm]\displaystyle
(U^1_1)^0_{\phantom{\mu}0}=-\frac{m^2}{2}\Big\{(\omega_1+\overline\omega_1)(W\Omega_1
A\,,W\Omega_1 A)-2\omega_1\overline\omega_1(W\Omega_1 A\,,\Omega_1
A)\Big\}-(MA)^0 (W\Omega_1A)_0\,.
\end{array}\end{equation}

The sign of $(T^0_1)^0_{\phantom{0}0}$ coincides with the sign of
$-\lambda_1$, see (\ref{3T00}). The linear combination of $(U^0_1)$
and $(U^1_1)$ does give rise to a positive conserved tensor unless
$\omega_1\overline\omega_1=0$ (Case B, $\lambda_2=0$). Thus, Cases B
and C of theory (\ref{EoM3}) are unstable unless the decomposition
(\ref{MExp}) has one simple nonzero root and double zero root.
Degrees of freedom of stable theory include one massive and one
massless vector mode.

\subsection{Case D} The polynomial (\ref{MExp}) has the simple real root $\lambda_1 $ multiplicity $3$, i.e.,
$$
M=m^2(W-\lambda_1 )^3=m^2(W^3-3\lambda_1 W^2+3\lambda_1
^2W-\lambda_1 ^3)\,.
$$
The comparison with (\ref{MExp}) brings us to the identification
$r=1$ and $s=0$. In this case, the index $i$ can take a single value
$i=1$, and $p_1=3$. The parametrization for the coefficients
$a_0,a_1,a_2$ of the polynomial (\ref{MExp}) reads
$$
a_2=-3\lambda_1 \,,\qquad a_1=3\lambda_1 ^2\,,\qquad a_0=-\lambda_1
^3\,.
$$

The general solution to the theory (\ref{EoM3}) consists of one
component. The new variables (\ref{xzA}) are not introduced.

The conserved tensors are constructed by the general rule
(\ref{WCT}) and parameterized by the indices $i=1$ and $p=0,1,2$.
The expressions for the tensors have the form

\begin{equation}\label{TRC}\begin{array}{l}\displaystyle
(T^0_1)^\mu_{\phantom{\mu}\nu}=-\frac{m^2}{2}\Big\{2(w^2A)^\mu(WA)_\nu+2(WA)^\mu(w^2A)_\nu+2\lambda_1 (wA)^\mu(wA)_\nu-\\[3mm]\displaystyle
\qquad-\delta^\mu_{\phantom{\mu}\nu}\big[2(w^2A)^\alpha(WA)_\alpha+\lambda_1 (wA)^\alpha(wA)_\alpha\big]\Big\}-(MA)^\mu A_\nu\,,\\[3mm]\displaystyle
(T^1_1)^\mu_{\phantom{\mu}\nu}=-\frac{m^2}{2}\Big\{2(wWA)^\mu(wWA)_\nu+2\lambda_1
(w^2A)^\mu(WA)_\nu+2\lambda_1
(WA)^\mu(w^2A)_\nu-\\[3mm]\displaystyle
\qquad-\delta^{\mu}_{\phantom{\mu}\nu}\big[(wWA)^\alpha(wWA)_\alpha+2\lambda_1
(w^2A)^\alpha
(WA)_\alpha\big]\Big\}-(MA)^\mu(WA)_\nu\,,\\[3mm]\displaystyle
(T^2_1)^\mu_{\phantom{\mu}\nu}=-\frac{m^2}{2}\Big\{2\lambda_1(w^2A)^\mu
(w^2A)_\nu+4\lambda_1(wWA)^\mu(wWA)_\nu+2\lambda_1^2(w^2
A)^\mu(WA)_\nu+\\[3mm]\displaystyle\qquad+2\lambda_1^2(WA)^\mu(w^2
A)_\nu-2\lambda_1^3(wA)^\mu(wA)_\nu-\delta^\mu_{\phantom{\mu}\nu}\big[\lambda_1(w^2A)^\alpha(w^2A)_\alpha+
2\lambda_1(wWA)^\alpha\times\\[3mm]\displaystyle\qquad\times(wWA)_\alpha+
2\lambda_1^2(w^2A)^\alpha(WA)_\alpha-\lambda_1^3(wA)^\alpha(wA)_\alpha
\big]\Big\}-(MA)^\mu(W^2A)_\nu\,,
\end{array}\end{equation}
where the notation $w=W-\lambda_1 $ is used.

The $00$-components read
\begin{equation}\label{TRC00}\begin{array}{l}\displaystyle
(T^0_1)^0_{\phantom{0}0}=-\frac{m^2}{2}\Big\{2(WA\,,W^2A)-3\lambda_1 (WA\,,WA)+\lambda_1 ^3 (A\,,A)\Big\}-(MA)^0A_0\,,\\[3mm]\displaystyle
(T^1_1)^0_{\phantom{0}0}=-\frac{m^2}{2}\Big\{(W^2A\,,W^2A)-3\lambda_1 ^2(WA\,,WA)+2\lambda_1 ^3(WA\,,A)\Big\}-(MA)^0(WA)_0\,,\\[3mm]\displaystyle
(T^2_1)^0_{\phantom{0}0}=-\frac{m^2}{2}\lambda_1\Big\{3(W^2A\,,W^2A)-6\lambda_1
(W^2A\,,WA)+\lambda_1^2(WA\,,WA)+2\lambda_1
^2(W^2A\,,A)\Big\}-\\[3mm]\displaystyle\qquad-(MA)^0(W^2A)_0\,.
\end{array}\end{equation}
One can check that the quantities (\ref{TRC}) are not combined into
a positive tensor. This result also applies to the case $\lambda_1
=0$. The theory with root of multiplicity three has to be considered
as unstable anyway.

\section{An example of stable self-interactions}
As we have seen, some of the higher derivative extensions of the
Chern-Simons theory admit positive conserved tensors at free level.
In this section, we provide an example of interaction in the Case A
such that the theory still has positive conserved tensor and remains
therefore classically stable.
 The equations of motion read
\begin{equation}\label{Mint}
\mathcal{M}A\equiv m^2(W-\lambda_1)(W-\lambda_2)(W-\lambda_3)A-
U'(\xi^\alpha\xi_\alpha)\xi=0\,,
\qquad\xi=\sum_{i=1}^3\beta{}^0_i\Lambda_i A\,,
\end{equation}
where $U(s)$ can be any scalar function, $U'(s)=\frac{dU(s)}{ds}$
and $\beta^0_i$ are treated as the parameters of interactions. The
interaction could be constructed by the factorization method of  the
papers \cite{KLS2014EurPhysC,KL-RFJ2014} that ensures survival of
selected conservation law of free theory at interacting level. Here,
we do not elaborate on the procedure for constructing the
interaction, we just examine consistency and stability of
interacting model.

 The theory admits the conserved tensor
\begin{equation}\label{CLint}
T^\mu_{\phantom{\mu}\nu}(A)=\sum_{i=1}^3\beta^0_i
(T^0_i)^\mu_{\phantom{\nu}\nu}(\Lambda_i
A)+\frac{1}{2}\delta^\mu_{\phantom{\mu}\nu}U(\xi^\alpha\xi_\alpha)\,,\qquad
\partial_\mu T^\mu_{\phantom{\mu}\nu}=-\partial_\nu\xi^\alpha
(\mathcal{M}A)_\alpha\,.
\end{equation}
With account of equations of motion it can be rewritten as
\begin{equation}\label{CLint2}
T^\mu_{\phantom{\mu}\nu}=-\sum_{i=1}^3\frac{m^2\beta{}^0_i\lambda_i}{2}\Big\{2(\Lambda_i
A)^\mu(\Lambda_i A)_\nu-\delta^\mu_{\phantom{\mu}\nu}(\Lambda_i
A)^\alpha (\Lambda_i
A)_\alpha\Big\}-U'(\xi^\alpha\xi_\alpha)\xi^\mu\xi_\nu+\frac{1}{2}\delta^\mu_{\phantom{\mu}\nu}U(\xi^\alpha\xi_\alpha)
-(\mathcal{M}A)^\mu\xi_\nu\,.
\end{equation}
The conserved tensor is positive if $\beta^0_i\lambda_i<0$ and
$U>0,U'<0$. The latter property is not satisfied by the polynomial
interactions. The admissible choice can be
$U(s)=\pi/2-\mathrm{arctg}(s)$, for example.

The consistent inclusion of interactions should not change the
degree of freedom number. The interaction (\ref{Mint}) is
consistent. This fact can be seen from decomposition of solution
into components (\ref{xi3}). The equations of motion for the
components take the form
\begin{equation}\label{E}
\mathcal{M}_i\xi_i\equiv m^2(W-\lambda_i)\xi_i-
U'(\xi^\alpha\xi_\alpha)\xi=0\,,\qquad
\xi=\sum_{j=1}^3\beta{}^0_j\xi_j\,,\qquad i=1,2,3.
\end{equation}
At the free level, elimination of longitudinal degree of freedom is
ensured by the transversality conditions
$\partial^\alpha(\xi_i)_\alpha=0\,,i=1,2,3.$ In non-linear theory,
the transversality conditions are modified but still remain the
first-order constraints,
\begin{equation}\label{dE}
\partial^\mu
(\mathcal{M}_i\xi_i)_\mu=\partial^\mu\left\{m^2\lambda_i(\xi_i)_\mu+
U'(\xi^\alpha\xi_\alpha)\xi_\mu\right\}=0\,.
\end{equation}
The degree of freedom number can be also covariantly computed
without depressing the order, e.g. by bringing the original higher
derivative equations into the involutive form as is it explained in
\cite{KLS-JHEP13}. Anyway, the equations (\ref{E}) still describe
three degrees of freedom, so the interaction (\ref{Mint}) is stable
(if $U>0, U'<0$) and consistent.

Rare examples are known of stable interactions in the higher
derivative systems. The best known example is $f(R)$-gravity
\cite{Strom,FaNa} where the canonical energy is bounded at
linearized level. This exceptional phenomenon happens because the
theory is strongly constrained. In the paper \cite{Pavsic}, the
stability of some interactions is demonstrated for the
Pais-Uhlenbeck oscillator (whose canonical energy is unbounded) by
numerical simulations. The stable interactions were recently
proposed for the Podolsky electrodynamics \cite{KLS2014EurPhysC} and
for the higher order Pais-Uhlenbeck oscillator \cite{KL-RFJ2014}.
The example of this section extends the limited list of known stable
interactions in higher derivative models.

 \vspace{0.2 cm}

\section*{Concluding remarks}
Let us summarize the results. In this paper, we suggest a simple
general procedure of constructing a family of higher order
symmetries and related conservation laws for the derived theories
whose equations are polynomial in certain operator (\ref{Wn}). For
the higher order extensions of the Chern-Simons theory (\ref{S}),
being an example of derived theory, we explicitly deduce the
conserved tensors. In some cases, depending on the structure of
roots in the polynomial (\ref{MExp}), the positive tensors exist
among the conserved quantities, while in the other cases, none of
the conserved quantities is positive. Once a positive conserved
tensor exists, the theory is classically stable, even though the
canonical energy is unbounded. In the third order examples of the
theory (\ref{S}) we notice that the stable theories realize the
irreducible unitary representations of the Poincar\'e group, while
the models admitting only unbounded conserved tensors correspond to
non-unitary representations. We also demonstrate that stable free
theory can admit consistent interactions that does not break the
stability.

Finally, we make remarks on stability at quantum level. Let us
mention that derived theories (\ref{Wn}) admit non-trivial Lagrange
anchors that can be constructed as polynomials in $W$ of order lower
than $n$. The construction of the anchor for the case $n=2$
demonstrated in \cite{KLS2014EurPhysC}. This can allow one to
quantize classically stable theory without loss of stability. As
established in \cite{KLS-JGP13,S-IJMPA14,S-IJMPA15}, every Lagrange
anchor leads to a Poisson bracket and Hamiltonian in the first order
formalism. The inequivalent Lagrange anchors lead to the canonically
inequivalent Poisson brackets, so the theory will be
multi-Hamiltonian in the first order formulation once it admits
different Lagrange anchors. As demonstrated in \cite{KLS-JMP10}, the
Lagrange anchor maps conservation laws to symmetries. In the
examples of classically stable higher derivative systems admitting
the different Lagrange anchors \cite{KLS2014EurPhysC,KL-RFJ2014},
the anchor exists such that maps the positive conserved quantity to
the time shift. This means that in the corresponding Hamiltonian
formalism (which is not unique, once there exist inequivalent
Lagrange anchors) the positive quantity can serve as Hamiltonian. As
example, let us mention that for the Pais-Uhlenbeck oscillator
positive Hamiltonians are known \cite{BOLONEK2005,DAMASKINSKY2006},
also at interacting level \cite{KL-RFJ2014}. As the Hamiltonian is
bounded at the classical level, we can hope to have bounded spectrum
of energy in quantum theory. In view of these reasons, we may expect
that classically stable higher derivative extensions of the
Chern-Simons model can remain stable at the quantum level once
appropriate Lagrange anchor is applied to quantize the theory.

\vspace{0.1 cm}

\noindent
 \textbf{Acknowledgments}. The authors thank K.~B.~Alkalaev and A.~A.~Sharapov for useful
discussions.

\noindent The work is partially supported by the Tomsk State
University Competitiveness Improvement Program.  DSC is partially by
the RFBR Grant 13-02-00551. SLL is partially supported by the RFBR
Grant 14-01-00489.

\end{document}